\documentclass[12pt]{article} 
\usepackage{geometry}
\usepackage{caption}
\usepackage{array}

\usepackage{chngcntr}
\counterwithout{equation}{section}

\newcommand{\Team}{2229048} 

\usepackage{wrapfig}

\usepackage{csvsimple}
\usepackage{enumitem}
\usepackage{amsmath,amssymb,amsthm,amsfonts}

\usepackage{hyperref}
\hypersetup{
    colorlinks = true,
    citecolor=black,
    filecolor=black,
    linkcolor=black,
    urlcolor=blue,
    }
\usepackage{lastpage}

\usepackage{booktabs} 

\usepackage{siunitx}

\usepackage{adjustbox}

\usepackage{xcolor}
\usepackage{fancyhdr}
\usepackage{graphicx} 

\allowdisplaybreaks

\usepackage{changepage}

\usepackage[vlined,ruled,linesnumbered]{algorithm2e}
\usepackage{tikz}
\usetikzlibrary{decorations.pathmorphing}

\title{Pedaling, Fast and Slow \\\textit{\large The Race Towards an Optimized Power Strategy}}
\author{Steven DiSilvio, Anthony Ozerov, Leon Zhou}
\date{February 21, 2022}

\begin{document}

\graphicspath{{.}}  
\DeclareGraphicsExtensions{.pdf, .jpg, .tif, .png}


\maketitle

\begin{abstract}
With the advent of power-meters allowing cyclists to precisely track their power outputs throughout the duration of a race, devising optimal power output strategies for races has become increasingly important in competitive cycling. To do so, the track, weather, and individual cyclist's abilities must all be considered. We propose differential equation models of fatigue and kinematics to simulate the performance of such strategies, and an innovative optimization algorithm to find the optimal strategy. 

Our model for fatigue translates a cyclist’s power curve (obtained by fitting the Omni-Power Duration Model to power curve data) into a differential equation to capture which power output strategies are feasible.  Our kinematics model calculates the forces on the rider, and with power output models the cyclist's velocity and position via a system of differential equations. Using fine-grained track data, including the slope of the track and velocity of the wind, the model accurately computes race times given a power output strategy on the exact track being raced.

To make power strategy optimization computationally tractable, we split the track into segments based on changes in slope and discretize the power output levels. As the space of possible strategies is large, we vectorize the differential equation model for efficient numerical integration of many simulations at once and develop a parallelized Tree Exploration with Monte-Carlo Evaluation algorithm. The algorithm is efficient, running in $O(ab\sqrt{n})$ time and $O(n)$ space where $n$ is the number of simulations done for each choice, $a$ is the number of segments, and $b$ is the number of discrete power output levels.

We present results of this optimization for several different tracks and athletes. As an example, the model's time for Filippo Ganna in Tokyo 2020 differs from his real time by just 18\%, supporting our model’s efficacy.
\end{abstract}

\newpage



\tableofcontents 
\newpage


\section{Introduction}

With the advent of power-meters allowing cyclists to precisely track their power outputs throughout the duration of a race, devising optimal power output distributions for races has become increasingly important in the world of competitive cycling \cite{sundstrom}. However, doing so involves considering a wide variety of factors, ranging from the biological limitations of the particular competitor to the topographical layout of the track the race will take place on. In particular, as competitions have gotten more advanced riders have become specialized for certain disciplines, such as sprinting and ultra-marathons, leading to drastic differences in rider capabilities which all need to be accounted for. Additionally, confounding factors such as inevitable deviations in power output by cyclists from specific power routines and weather variations on the day of the competition must also be considered when designing any routine. An ideal power output distribution would minimize the time necessary to complete a race subject to the constraint that the rider can feasibly carry out such a routine to its completion.

\section{Problem Statement and Assumptions}
Given the power curve of a cyclist, which for any length of time gives the maximum amount of constant power output they can sustain, as well as track-specific data, we are tasked with devising a model to output an optimal power distribution for completing a race in minimal time.  In addition, we analyze the effect of deviation from the optimal strategy on race time, and allow for varying weather conditions to be accounted for in the model.  In doing so, we make the following assumptions:
\begin{itemize}
    \item A rider's power curve is monotonically decreasing, and in particular is a bijection from $\mathbb{R}$ to a closed subinterval of $\mathbb{R}^+$. This is reasonable, as it would not be sensible if, for example, someone's maximum sustainable power output for 10 seconds is 1000W, and for 9 seconds is 800W. This assumption is useful as it means the power curve is invertible, and a unique duration can be computed given a power output level.
    \item Cyclists are able to recover energy levels as they ride by cycling with less power.  In particular, each cyclist has a critical power output at which they neither gain or lose energy, and which they could theoretically maintain indefinitely.  Cycling above this critical power will fatigue a cyclist over time, and cycling below it will allow them to regain energy.  This assumption is supported by the literature, and has been used in several other studies \cite{physics}.  
    \item It is not feasible for a cyclist to have a power distribution plan in which they must continuously vary their power, as doing so would be far too complex for a rider to follow.  Instead, the race must effectively be broken up into discrete segments based on the distance a rider has covered, and during such a segment a rider will be instructed to maintain a specific constant power level.
    \item A cyclist can effectively be viewed as traveling in one dimension, having a velocity at any given instant which translates to a speed along the direction of the path.  Dimensional considerations such as varying elevations lead to the exertion of forces on the rider, but these forces can be broken into components parallel to the cyclist’s direction of travel (which will affect their speed) and perpendicular to it (which will impose restrictions on the velocity attainable to still remain along the course).

\end{itemize}

\section{Modeling Methodology}


We obtain track data in the form of GPX traces from \textit{La Flamme Rouge} \cite{trackdata}, a European provider of cycling route data. This data contains latitude, longitude, and elevation, and matches remarkably well with officially-published race profile images.
\subsection{Track Data}
\begin{figure}
    \centering
    \includegraphics[width=0.5\textwidth]{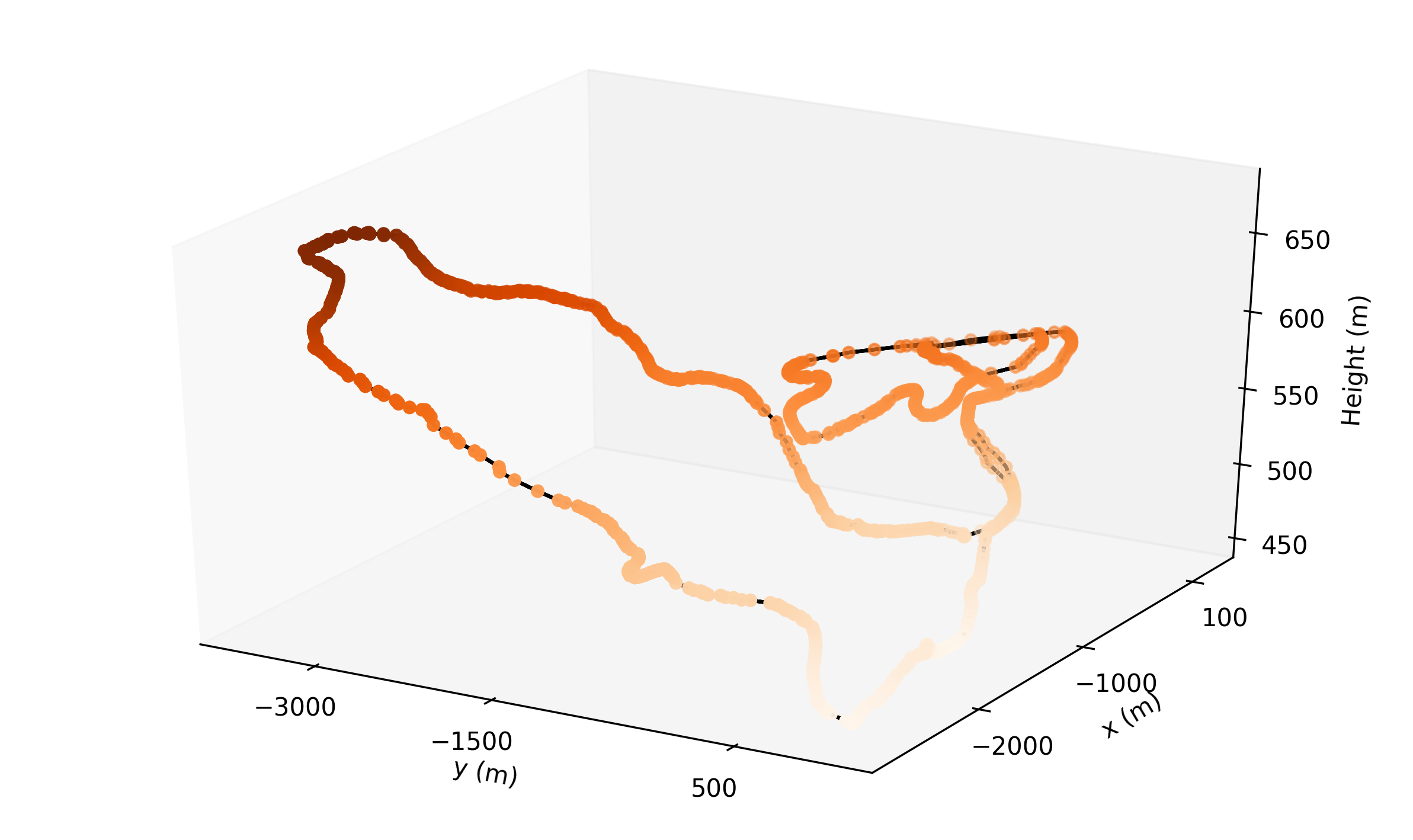}
    \caption{Tokyo 2020 track data visualized with local East (y), North (x), and Up (Height) coordinates in the tangent plane at the track start point. Elevation exaggerated.}
    \label{fig:3D-Track}
\end{figure}
\paragraph{Distances} Using this data, we calculate the cumulative distance $d_i$ (on the horizontal plane) at each point $i$ along the track according to:
\[d_i = \sum_{j=1}^{i-1} d_{j,j+1},\]
where $d_{j,j+1}$ is the distance between points $j$ and $j+1$. We calculate $d_{j,j+1}$ according to the haversine formula \cite{haversine}, which is used for computing the lengths of great circles based on the latitude and longitude of the two endpoints:
\[d_{ij}=2r\arcsin\left(\sin^2\left(\frac{\lambda_j-\lambda_i}{2}\right)+\cos(\lambda_i)\cos(\lambda_j)\sin^2\left(\frac{\mu_j-\mu_i}{2}\right)\right),\]
where $\lambda_i,\lambda_j$ are the latitudes of points $i$ and $j$ and $\mu_i,\mu_j$ are the longitudes of points $i$ and $j$.
\paragraph{Slopes}
If we let $a_i$ be the elevation at point $i$, also calculate the slope angle $\theta_i$ at each point according to:
\[\theta_i = \arctan\left(\frac{a_{i+k}-a_{i-k}}{d_{i+k}-d_{i-k}}\right)\]
where $k$ is a smoothing factor, which we set to 5. This smoothing is necessary as the elevation data is at one-meter resolution.

\paragraph{Curvature}
One key track statistic is its curvature. As is explained in the section below, the curvature metric required is the radius of the circle that the curve is following. To determine this, we fit a circle to a small set of $k$ points around every point, as using three points to get the unique circle would be less robust. We do this by solving the following least-squares problem:
\[\min_{x_c,y_c,r}\sum_{i=1}^k \left|r-\sqrt{(x_i-x_c)^2+(y_i-y_c)^2}\right|.\;\;\cite{circles}\]
Effectively, given the set of points with coordinates $(x_1,y_1),\ldots,(x_k,y_k)$, we find the coordinates $(x_c,y_c)$ and radius $r$ that minimize the distance from the points to the circle. Note that to get a radius in meters, the latitudes and longitudes of the points are converted to coordinates in meters in the local tangent plane at the start point of the track.

\begin{figure}[!h]
\centering
\includegraphics[width=0.8\textwidth]{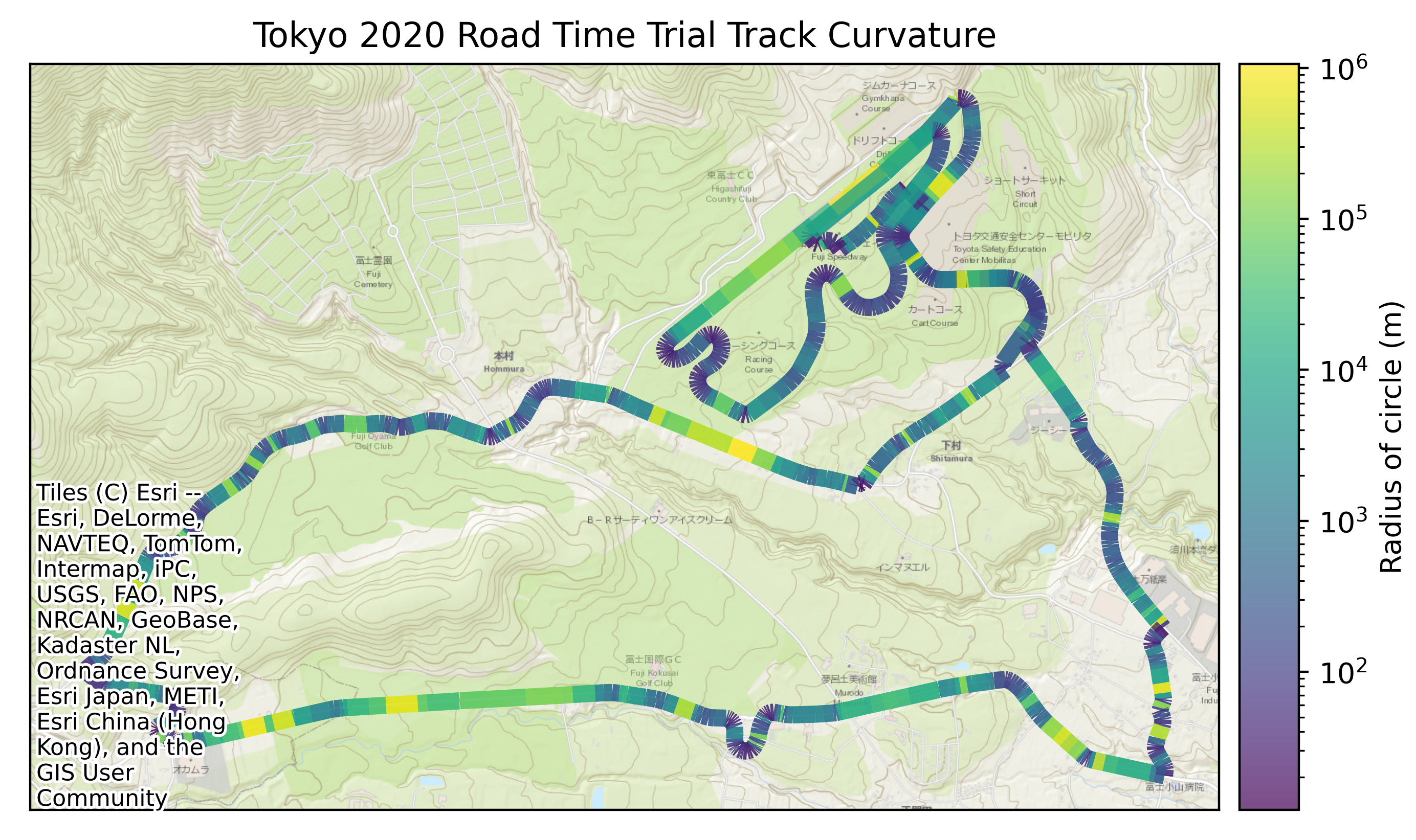}
\caption{Radii of fitted circles on the Tokyo 2020 Men's Road time trial route.} \label{fig:radius}
\end{figure}

\paragraph{Heading}
To account for the effect of wind, we must know the velocity of the wind in the direction of the cyclist. This depends solely on the windspeed and the angle of the wind relative to their direction of motion. Thus, the direction of motion at every point along the route must be calculated. With $\lambda_i,\lambda_{i+1}$ and $\mu_i, \mu_{i+1}$ we may use the following equation to calculate the heading $\gamma_i$ at every point, using its coordinates and those of the next point: 
\[x=\sin(\mu_{i+1}-\mu_i)\cos(\lambda_{i+1})\]
\[y=\cos(\lambda_i)\sin(\lambda_{i+1})-\sin(\lambda_i)\cos(\lambda_{i+1})\cos(\mu_{i+1}-\mu_i)\]
\[\gamma_i = \text{atan2}(x,y). \;\;\cite{heading}\]

\paragraph{Interpolation and gridding} The above computations give track data at every point in the GPX trace. However, the models below require track data at arbitrary points. To account for this, we pre-calculate interpolated track data on a 10m grid of points for efficient access during the simulations. At each gridpoint $x$, a data point $z(x)$ is calculated according to a linear interpolation between the two nearest points $i$ and $i+1$:
\[w(x) = \frac{d_{i+1}-x}{d_{i+1}-d_i}\]
\[z(x) = w(x)z_i + (1-w(x))z_{i+1} \]

\subsection{Kinematic Differential Equation Model}

\paragraph{Motivation}
Differential equation models lend themselves very naturally to the study of kinematics.  In particular, knowing a cyclist's power output over the entire duration of a race, as well as the topography of the track itself, it is possible to set up differential equations modeling incremental changes in their speed and position, allowing for the time they take to complete a race to be numerically solved.  

\paragraph{Preliminary assumptions} In order to simplify our model to be computationally feasible, the following reasonable assumptions were made:
\begin{itemize}
    \item Riders can effectively be viewed as traveling in one dimension (i.e. along the direction of the track), with all imposed forces able to be broken up into components parallel and perpendicular to this direction of motion 
    \item Turning corners does not affect a rider's speed along their direction of motion, but instead imposes a limitation on the maximum speed attainable in order to stay along the curve (this is in line with the fact that centripetal force is always perpendicular to the velocity of a moving object, and hence imposes no work)
    \item Power outputs from power curves are the effective power that one is able to produce, i.e. the power driving them forward, and hence internal power loss (due to gear and chain friction) are not considered
\end{itemize}

\begin{table}[!ht]
\centering
\caption{Differential equation model variables and constants.}
\begin{tabular}[t]{cp{2.0in}lcp{1.7in}}
\hline
Variable &\multicolumn{3}{l}{Description}&Units\\
\hline
$x$&\multicolumn{3}{l}{Distance covered along the track}&m\\
$x_h$&\multicolumn{3}{l}{Horizontal distance covered}&m\\
$v$&\multicolumn{3}{l}{Velocity}&m/s\\
$a$&\multicolumn{3}{l}{Acceleration}&m/s$^{2}$\\
$\theta$&\multicolumn{3}{l}{Angle of incline}&radians\\
$P$&\multicolumn{3}{l}{Power output}&W\\
$\gamma$&\multicolumn{3}{l}{Cyclist heading}&radians\\
$v_w$&\multicolumn{3}{l}{Windspeed}&radians\\
$\gamma_a$&\multicolumn{3}{l}{Wind heading}&radians\\
$v_a$&\multicolumn{3}{l}{Effective airspeed against cyclist's motion}&m/s\\
$F$ & \multicolumn{3}{l}{Net force parallel to cyclist's motion} & N \\

\hline
Constant &Description&Value&Units&Source/Rationale\\
\hline
$A$&Frontal area of cyclist&$.4$&m$^2$& \cite{area} \\
$m$&Mass of cyclist & [various] & kg & \cite{stats}\\
$g$& Gravitational acceleration &9.81& m/s$^2$& \\
$C_d$& Coefficient of drag& 0.6 & - & \cite{dragcoefficient}  \\
$\mu_s$ & Static friction coefficient & [various] & -& \cite{static} \\
$\mu_r$ & Rolling resistance & .0025 & - & \cite{rolling}\\
\hline
\end{tabular}

\label{table:DEt}
\end{table}

\paragraph{Horizontal and Track Distance} At any given instance, where $x(t), x_h(t)$ are the distances traversed along the track and horizontally respectively as a function of time $t$, we have that:
\[\frac{dx_h}{dt} = \frac{dx}{dt} \cos\big( \theta \left(x_h\right) \big)\]
This makes intuitive sense, as assuming a constant angle $\theta$ of inclination for some stretch of track, by elementary trigonometry clearly $\Delta x_h = \Delta x \cos (\theta)$.  Instantaneously at a given time $t$, the angle $\theta(x_h(t))$ is computable via the method described in the previous section.

\paragraph{Air speed} Given a cyclist's velocity $v$ and heading $\gamma$, as well as the speed $v_w$ and heading $\gamma_w$ of the wind, we calculate the effective speed of the air moving against their motion,  $v_a$ , as follows:
\[v_a = v - v_w\cos (\gamma - \gamma_w)\]
In particular, under conditions with no wind (i.e. $v_w = 0$) we have that $v_a = v$, as the speed of the air moving towards the rider is simply the speed with which they are traveling through the medium.  

\paragraph{Force parallel to cyclist's motion}

\begin{figure}
    \centering
    
    \includegraphics[width=0.6\textwidth]{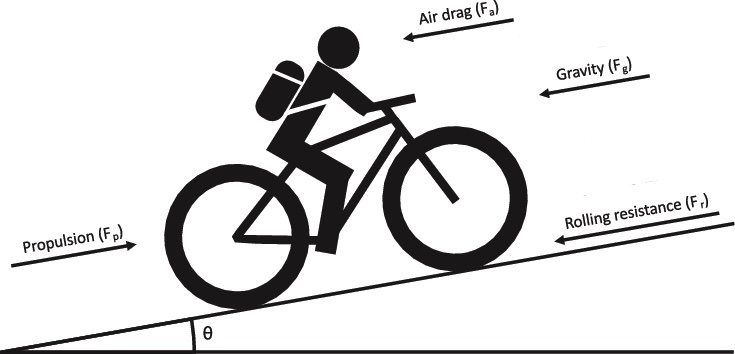}
    \caption{Free body diagram of forces parallel to cyclist's direction of motion. Adapted from \cite{FreeBody}}
    \label{fig:FreeBody}
    
\end{figure}

Forces acting parallel to the cyclist's motion under consideration are those due to gravity, rolling resistance, drag and their own power output (i.e. forward propulsion) as seen in figure \ref{fig:FreeBody}.  Where $P$ is the power function taking distance along the track as input which we wish to optimize for, the forces are given as follows \cite{physics}:

\begin{align*}
F_a &= \frac{1}{2}C_dAv_a^2 \\
F_g &= mg\sin \big(\theta (x_h) \big)\\
F_r &= \mu_r mg\cos \big(\theta (x_h) \big)\\
F_p &= \frac{P(x)}{v}
\end{align*}
Thus, where each of $F_p, F_a, F_g$ and $F_r$ are positive magnitudes, we have that the total force acting on the cyclist is given by:
\[F = F_p - F_a - F_g - F_r\]
where the positive direction is the direction of the cyclist's travel.

\paragraph{Forward acceleration}
Applying Newton's Second Law of Motion and substituting the forces as above, we immediately get that:
\begin{equation*}\label{eq:dv_dt}
\frac{dv}{dt}= \frac{F}{m}
\end{equation*}

\paragraph{Maximum velocity around curves} While traveling around a banked curve, forces due to static friction and gravity act perpendicular to the motion of the cyclist to change their direction to keep them along the curve while not affecting the magnitude of their velocity.  However, given a portion of track with bank angle $\beta$ and curvature radius $r$, where $r$ corresponds to the radius of the circle fitted via the least squares method described above, the maximum attainable velocity is given by:
\[v_{max} = \sqrt{\frac{rg(\sin(\beta) + \mu_s \cos(\beta))}{\cos (\beta) - \mu_s \sin (\beta)}} \cite{bank}\]
Traveling above $v_{max}$ would cause the rider to fly out of the curve, and hence we cap their velocity at $v_{max}$ at any given point along the track as we assume a cyclist would always break to assure they stay on the path.

\subsection{Omni-Power Duration Model for Cyclists}

\paragraph{Model} We calculate power curves of various athletes using the Omni-PD model \cite{Omni-PD} on their public data. The model has been previously shown to be accurate in modeling both high-intensity and endurance cycling. These athletes ranged from amateurs to Olympians that have competed at the 2020 Tokyo time trials. Data was pulled from Strava, and we took their maximal mean power levels at various set intervals. This is the maximum energy they outputted within any interval divided by the duration of that interval. Usually riders are able to generate enormous power for seconds, but the power levels quickly taper out due to the body building up too much lactic acid and will switch from anaerobic to aerobic mechanisms. At longer intervals of exercise, cyclists enter into a steady state, which is a relatively flat energy level. 

All of this is accounted for by the model. Using non-linear least squares, we can fit the curve to those points as seen in figure 4.

\paragraph{Deriving Critical Power and Work Capacity}
By tuning these parameters to fit the power levels at certain intervals, we can also deduce constants for each individual, such as their Critical Power threshold $(P_c)$, and their Anaerobic Work Capacity $(W')$. Critical power is the maximum sustainable energy over an extended period of time. Usually for an athlete this will be around 30 minutes though it varies by individuals \cite{Omni-PD}. Any power level above Critical power will drain from the work capacity, and cause fatigue when the entire energy capacity is depleted \cite{WorkFatigue}. On the other hand, going below that threshold will allow riders to recover their energy slightly.

\begin{table}[!ht]
\centering
\caption{Power Curve variables and constants.}
\begin{tabular}[t]{cp{1.8in}lcp{1.9in}}
\hline
Variable &\multicolumn{3}{l}{Description}&Units\\
\hline

$P_{max}$&\multicolumn{3}{l}{Max Power}&W\\
$P_C$&\multicolumn{3}{l}{Critical Power}&W\\
$W'$&\multicolumn{3}{l}{Work above $P_C$ (Anaerobic Work Capacity)}&W\\
$t$&\multicolumn{3}{l}{Time}&s\\
$T_{cpmax}$&\multicolumn{3}{l}{Time sustained at $P_C$}&s\\
\hline
Constant &Description\\
\hline
$\beta$&Linear Constant & -  \\

\end{tabular}

\label{table:Omni-Var}
\end{table}

\begin{equation}
f(t)=
    \begin{cases} \label{eq:Omni-PD}
      \frac{W'}{t}*(1-e^{t*\frac{P_{max}-P_C}{W'}}) + P_C & t\leq T_{cpmax} \\
      \\
      \frac{W'}{t}*(1-e^{t*\frac{P_{max}-P_c}{W'}}) + P_C -\beta*ln(\frac{t}{T_{cpmax}}) & t\geq T_{cpmax} 
   \end{cases}
\end{equation}

For each power level, we can take the inverse of the function to determine the maximum time each athlete can sustain a certain power level for. These power levels will become the choices that a rider can make at each segment of the race. 

\paragraph{Professional Data}
All our simulations are based off the power curves we computed for three professional cyclists shown in table \ref{table:Profiles}. Their MMP were derived from power recordings they achieved in either races or high-intensity practices. 
\begin{table}[!ht]
\centering
\caption{Rider Profile and Information \cite{stats}}
\begin{tabular}[t]{cp{1.8in}lcp{1.9in}}
\hline
Name &\multicolumn{1}{l}{Racer Type}&Gender&Weight(kg)\\
\hline

    Filippo Ganna & \multicolumn{1}{l}{Time Trial Specialist}&M&82\\
    Mathieu van der Poel & Puncheur & M & 75\\
    Chloe Dygert & Time Trial Specialist & W & 67\\
\hline
\end{tabular}
\label{table:Profiles}
\end{table}

\begin{figure}
    \centering
    \includegraphics[width=0.5\textwidth]{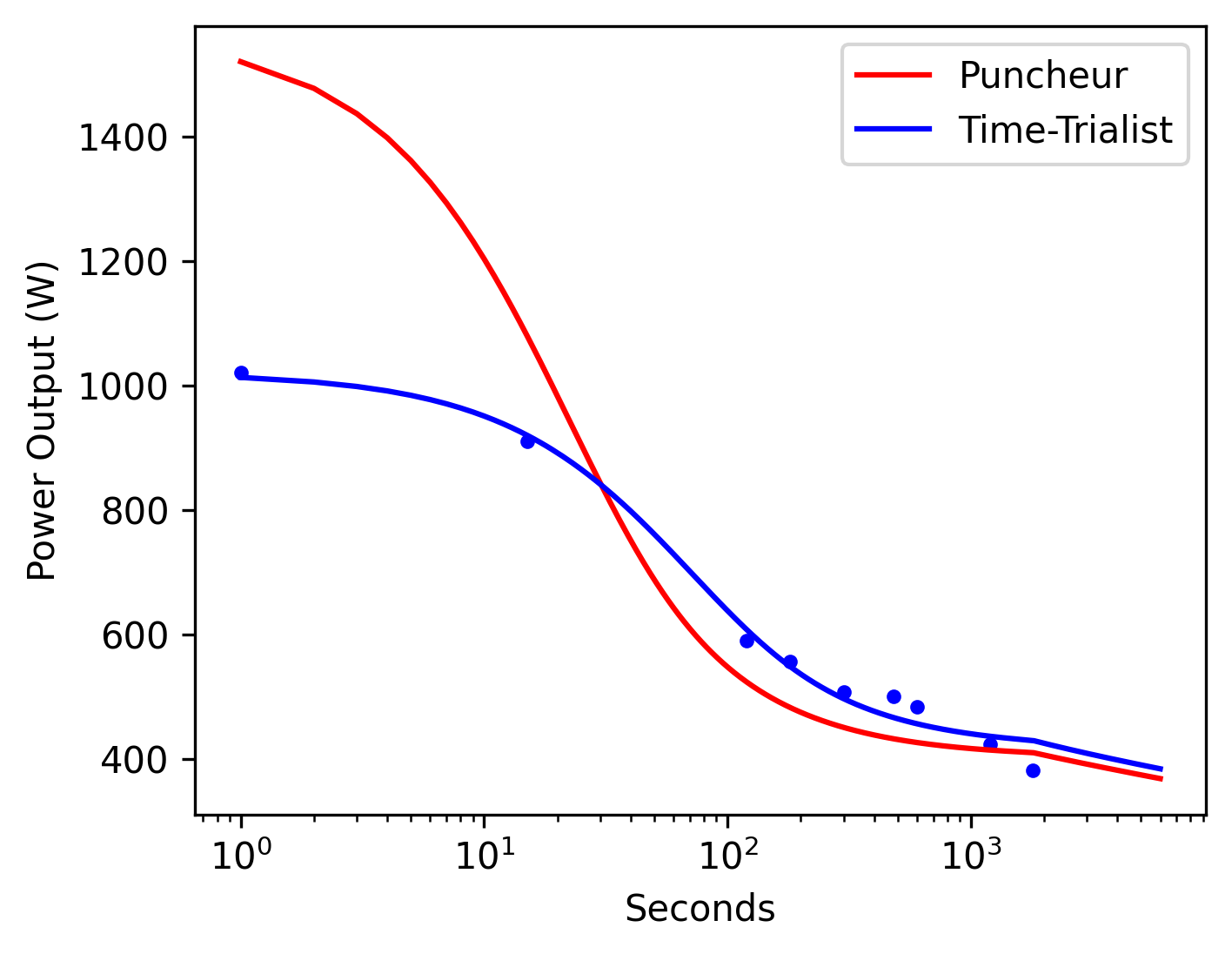}
    \caption{Comparison of two different cyclist profiles, as fitted by the Omni-PD model. The Puncheur, a cyclist who specializes in short steep hills, is modelled off the data for Mathieu van der Poel. The time trial specialist is modelled from Filippo Ganna's rides. 
         \cite{Filippo}\cite{VanDer}}
    \label{fig:Cycling Profile}
\end{figure}

\subsection{Fatigue Differential Equation Model}
A cyclist cannot maintain a high level of power output indefinitely, and hence developed a model for fatigue which tracks their energy level, constrained to the interval $[0,1]$ and initialized to $1$, throughout the duration of the race. As a cyclist's power curve represents how long they can maintain a given level of power output, we present the following simple model for fatigue, where $f^{-1}$ is the inverse of a given cyclist's power curve:

  \[
    \frac{dE}{dt}=
    \begin{cases}
        \frac{-1}{f^{-1} (P(x_h))} & \text{if } P(x_h) > P_C   \\
        \frac{1}{7200P_C}(P_C - P(x_h)) & \text{if } P(x_h) \leq P_C
    \end{cases}
    \]
    
    This model accounts for the fact that cyclist's can theoretically sustain their critical power indefinitely, while levels above this will diminish their energy and levels below this will allow them to recover energy.  In particular, given a power level above $P_C$, the less time a cyclist can maintain it based on their power curve the greater the rate of decrease in their energy levels which outputting it would cause.  The constant $\frac{1}{7200P_C}$ was chosen such that after 2 hours (7,200 seconds) of not riding (i.e. maintaining a power level of 0), a cyclist completely fatigued would have their full energy restored.  
    
    In regards to the race simulation, a cyclist can never allow their energy level to drop below 0, and hence this fatigue model provides a natural constraint on power output.  As a cyclist nears 0 energy, they are incentivized to output lower power levels to either conserve or regain energy.  Additionally, a cyclist's energy is capped at 1, and hence cycling below $P_C$ can not indefinitely increase a cyclist's energy.

\subsection{Optimization: Tree Exploration with Monte-Carlo Evaluation}
\begin{figure}
    \centering
    \includegraphics[width=0.5\textwidth]{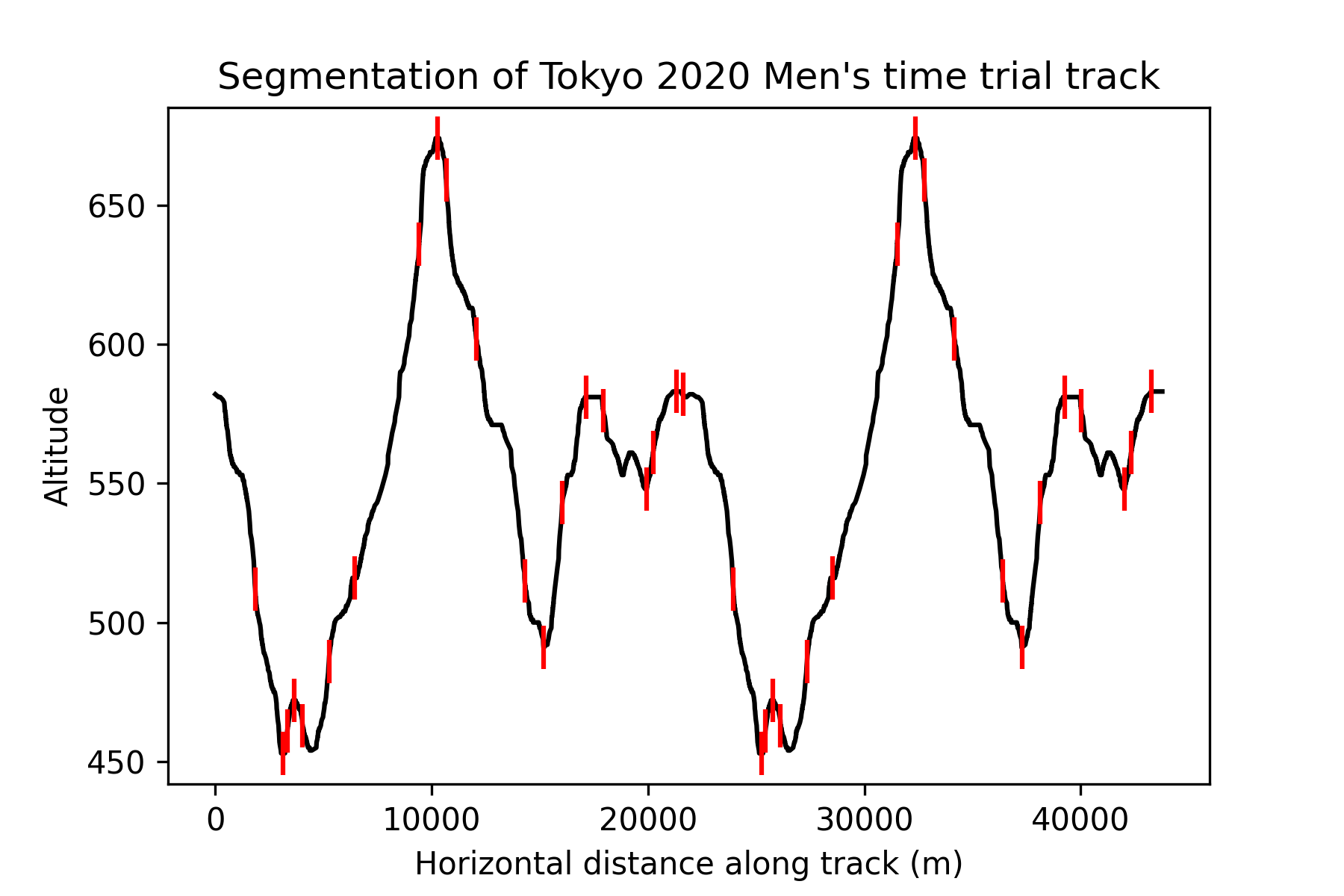}
    \caption{Segmentation of Tokyo 2020 Men's time trial track, done using slope and altitude data}
    \label{fig:tokyo_seg}
\end{figure}

With models for fatigue and cycling kinematics, we may now optimize the power strategy to minimize the time taken to complete the time trial.
\paragraph{Track segmentation}
The slope of the track is a dominant factor in the cycling kinematics equation. Indeed, studies of cycling power strategies often focus on the course profile \cite{sundstrom}. For this reason, we segment the track based on changes in slope. This is done by taking local minima and maxima of elevation (which are necessarily locations where the slope changes from negative to positive or positive to negative, respectively) and local minima and maxima of slope---thus we identify both changes in slope direction and slope extremes. Figure \ref{fig:tokyo_seg} displays the segmentation of the Tokyo route.

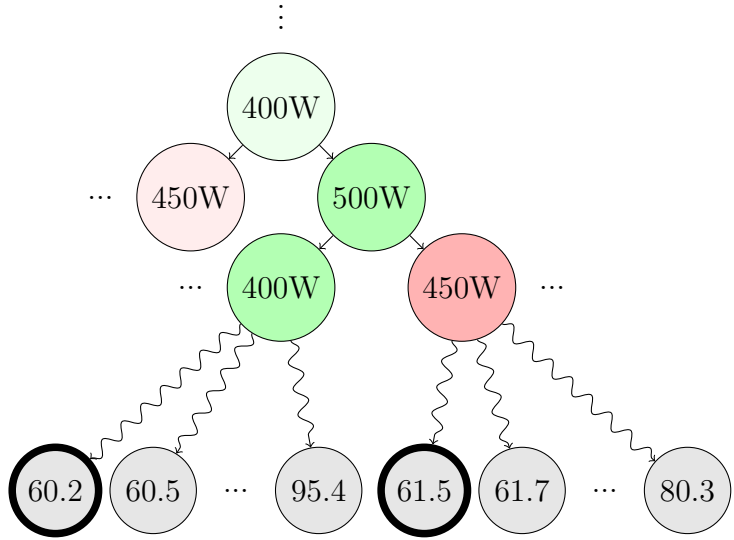
\begin{figure}
\centering
\tikzstyle{gnode} = [circle, minimum width=1cm, minimum height=1cm,text centered, draw=black]
\tikzstyle{child} = [circle, minimum width=1cm, minimum height=1cm,text centered, draw=black, fill=green!20]
\tikzstyle{leaf} = [circle, minimum width=1cm, minimum height=1cm,text centered, draw=black, fill=black!10]
\tikzstyle{dots} = [circle, minimum width=0cm, minimum height=0cm,text centered, draw=white!0, fill=green!0]
\tikzstyle{arrow} = [thick,->,>=stealth]
\begin{tikzpicture}[node distance=1.7cm, scale=1, decoration=snake]
\node[gnode, fill=green!7] (r2) {400W};
\node[dots] (d3) [above of=r2, yshift=-0.4cm] {\vdots};
\node[gnode, fill=red!7] (r3) [below left of =r2] {450W} edge[<-] (r2);
\node[dots] (d3) [left of=r3, xshift=0.5cm] {...};
\node[gnode, fill=green!30] (root) [below right of =r2] {500W} edge[<-] (r2);
\node[child, fill=green!30] (c1) [below left of=root] {400W} edge [<-] (root);
\node[child, fill=red!30] (c2) [below right of=root] {450W} edge [<-] (root);
\node[dots] (c3) [right of=c2, xshift=-0.5cm] {...};
\node[dots] (c3) [left of=c1, xshift=0.5cm] {...};
\node[leaf] (c1l1) [below of = c1, yshift=-1cm, xshift=0.5cm] {95.4};
\draw[->,decorate,decoration=snake] (c1) -- (c1l1);
\node[dots] (d1) [left of = c1l1, xshift=0.6cm] {...};
\node[leaf] (c1l2) [left of = d1, xshift=0.6cm] {60.5};
\draw[->,decorate,decoration=snake] (c1) -- (c1l2);
\node[leaf, line width=1mm] (c1l3) [left of = c1l2, xshift=0.4cm] {60.2};
\draw[->,decorate,decoration=snake] (c1) -- (c1l3);
\node[leaf, line width=1mm] (c2l1) [below of = c2, yshift=-1cm, xshift=-0.5cm] {61.5};
\draw[->,decorate,decoration=snake] (c2) -- (c2l1);
\node[leaf] (c2l2) [right of = c2l1, xshift=-0.4cm] {61.7};
\draw[->,decorate,decoration=snake] (c2) -- (c2l2);
\node[dots] (d2) [right of = c2l2, xshift=-0.6cm] {...};
\node[leaf] (c2l3) [right of = d2, xshift=-0.6cm] {80.3};
\draw[->,decorate,decoration=snake] (c2) -- (c2l3);
\end{tikzpicture}
\caption{Diagram of the optimization method. Having already chosen some sequence of power levels ($\ldots$, 400W, 500W) for a sequence of segments, we wish to decide between 400W, 450W, and some other power levels for the next segment. For each of these choices, we randomly pick $n$ assignments of power levels for the remaining segments, and calculate the time taken to cycle the route using each of the random assignments. For 400W, the best time was 60.2 minutes, whereas for 450W, it was 61.5 minutes. Thus we choose 400W as the power level for the next segment. We repeat until we have chosen a power level for each segment.}\label{fig:optimization}
\end{figure}

\paragraph{Optimization problem}
There are three factors which determine the most effective optimization method for our formulation of the problem:
\begin{enumerate}
    \item Very large search space. With even just 20 segments and 5 power levels, there are $5^{20}$ possible power strategies.
    \item The differential equation model can be vectorized to integrate an arbitrary number $n$ of power strategies simultaneously in $O(\sqrt{n})$.
    \item Due to momentum and fatigue levels, the effect on finishing time of choosing a power level for one segment depends on the choices for all of the other segments.
\end{enumerate}
\paragraph{Optimization method}
Given these properties, we design a stochastic optimization method which is linear in the number of segments and the number of power levels and converges to the optimal solution with sufficiently large $n$. Starting from the first segment, we repeatedly choose the power level for the next segment which coincides with that in the optimal solution (i.e. that which minimizes time). Since it is computationally intractable to test every possible power strategy, we may instead choose $n$ random power strategies for the rest of the segments and get their time results. We then use the minimum result of the random power strategies to select the power level for the next segment. As $n\to \infty$, the minimum result will be that of the optimal solution. Figure \ref{fig:optimization} illustrates this process. In practice, rather than evaluating choices using the minimum time, we use the $k$th lowest time. In the limit, this will still find the optimal solution, but with smaller $n$ this makes the optimization less sensitive to the stochasticity of the process.

\section{Model Results}

\subsection{Sample Solutions: Tokyo 2020 Time Trial}
\begin{figure}
    \centering
    \includegraphics[width=\textwidth]{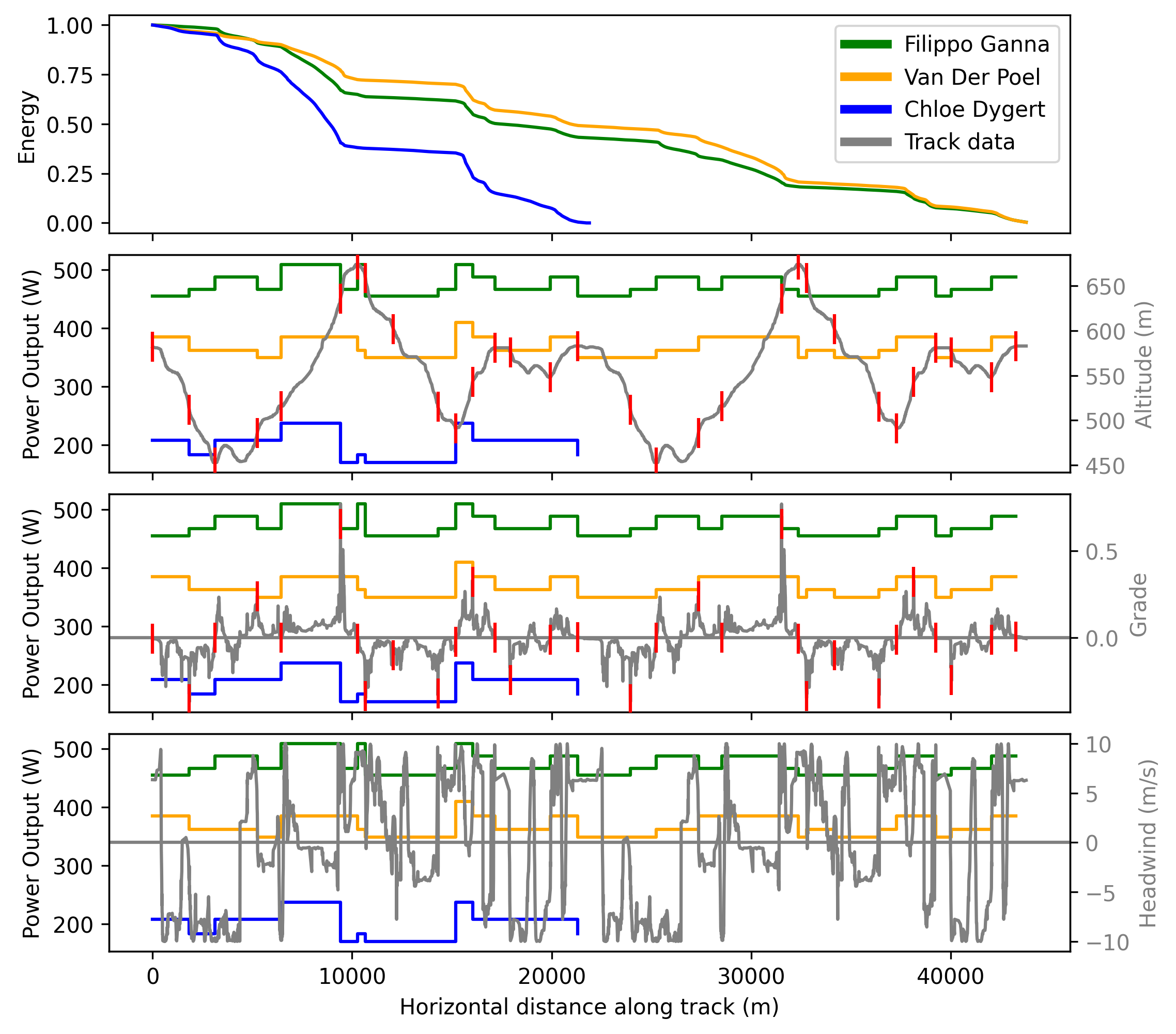}
    \caption{The optimal strategies for the three cyclists examined when optimized for the Tokyo 2020 time trial route. Note that the women's time trial route is half the distance. We see that all of the strategies tend to apply more power when going uphill and much less when going downhill. We also see that the strategies of Filippo Ganna and Van Der Poel do differ slightly. Ganna reserves bursts of power for major hills. Van Der Poel, a hill climber, applies more power over less-steep hills as well, e.g. he applies a high power level over the whole slope from 28,000m to 33,000m where Filippo does not.}
    \label{fig:tokyo-solution}
\end{figure}

We use an Euler approximation to numerically solve the Differential Equation Model for the three cyclists studied on the Tokyo 2020 Time Trial route. Figure \ref{fig:tokyo-solution} displays solutions for the Tokyo 2020 Time Trial for three different athletes.

\subsection{Sensitivity to Weather}
To determine the optimal strategy's sensitivity to deviations in weather, we may simulate the finishing times of the optimal strategy during different weather conditions than those optimized for.
\begin{figure}
    \centering
    \includegraphics[width=0.49\textwidth]{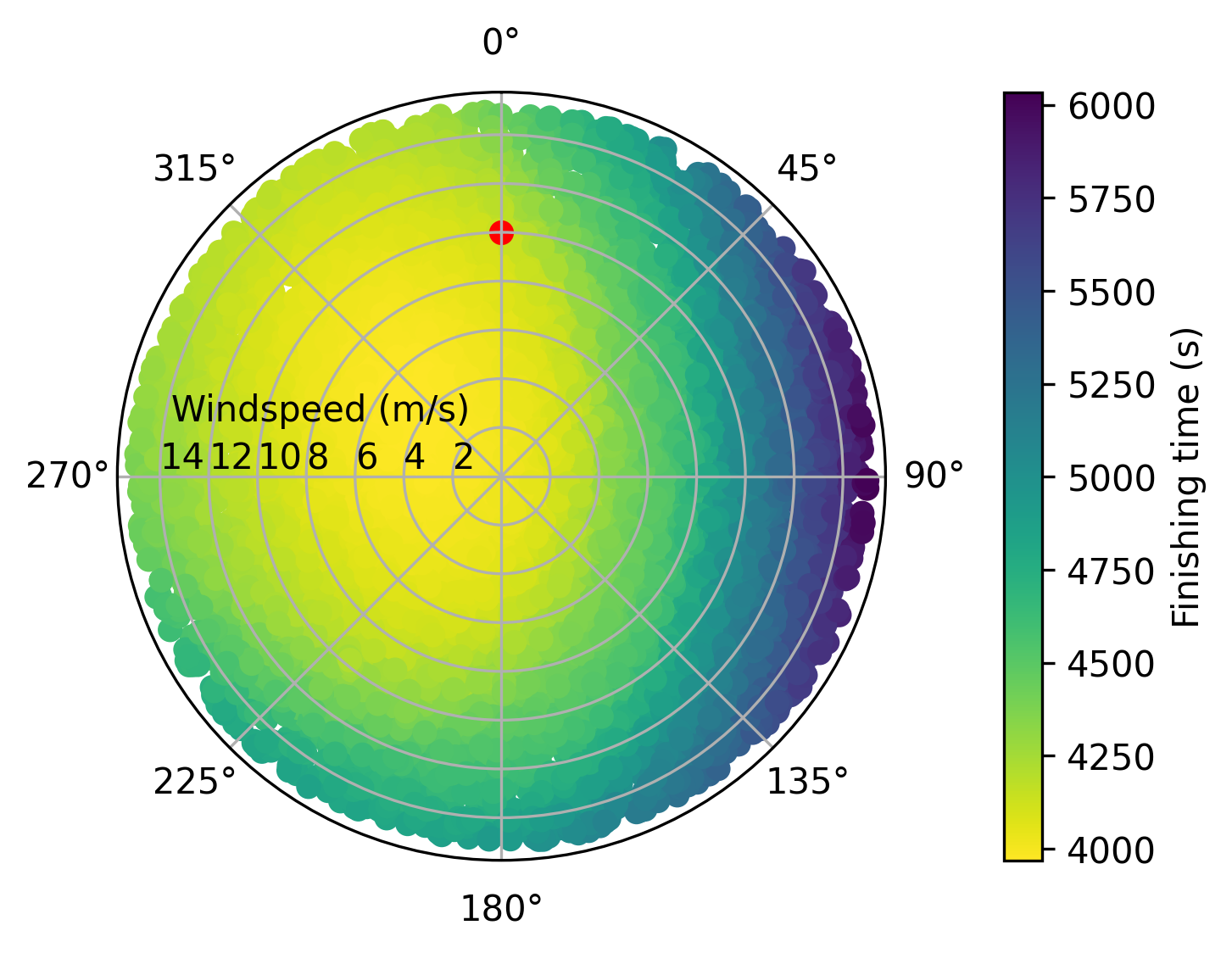}
    \includegraphics[width=0.49\textwidth]{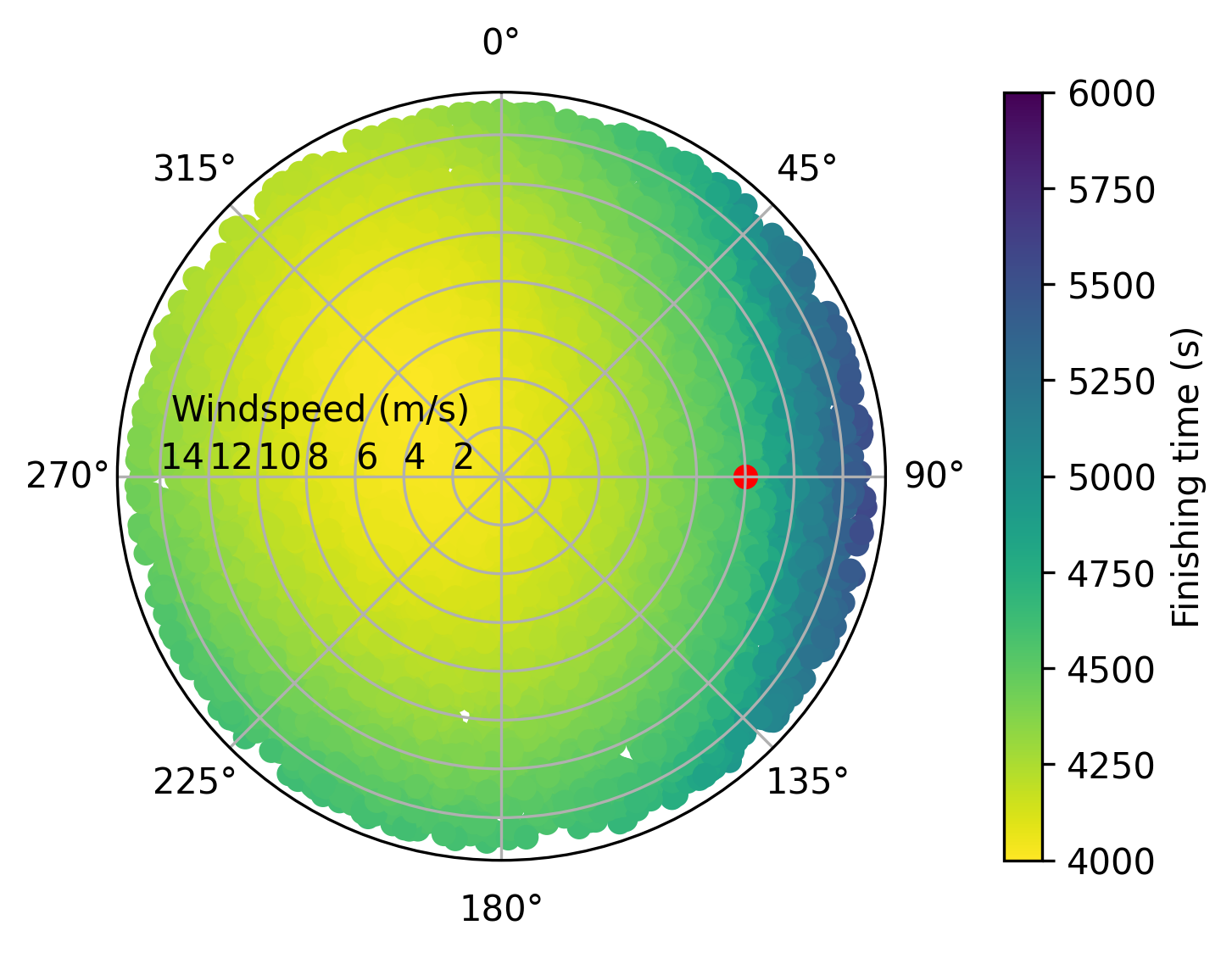}
    \caption{The left plot displays finishing times for a strategy for Filippo Ganna in Tokyo 2020 optimized for a northward wind with speed 10m/s (indicated by the red point) under a range of other windspeeds and directions. The strategy still performs well under winds to the West and North. The strategy performs significantly worse under winds blowing East. On the right, having optimized instead for an eastward wind with speed 10m/s, we see that the optimization does indeed take into account wind conditions, as much better times are achieved with eastward winds than when optimizing for 10m/s North.}
    \label{fig:wind-finish}
\end{figure}
\paragraph{Wind}
Ganna's finishing time with the optimal strategy for a northward 10m/s wind is 4140 seconds. We run 10,000 simulations of this strategy with wind direction chosen uniformly randomly and windspeed chosen uniformly randomly in the interval $[0,10]$. Figure \ref{fig:wind-finish} displays the finishing times of these simulations. We find that differences in windspeed and direction do affect the finishing times, but strategies (whose optimization is not based solely on windspeed), can still perform as well or better under a range of weather conditions.

\begin{table}
    \centering
    \caption{Filippo Ganna's time results (in seconds) under the optimal strategy in the Tokyo 2020 time trial under different weather conditions. Note that, when the weather conditions match those optimized for, the times are roughly the same. When the weather conditions do not match, times worsen significantly.}
    \begin{tabular}{ccc}
        \textbf{Optimized weather} &\textbf{No rain in simulation} & \textbf{Rain in simulation}\\\hline
        No rain & 4105 & 4315\\
        Rain & 4416 & 4099\\ 
    \end{tabular}
    \label{tab:rain}
\end{table}

\paragraph{Rain}
Previous results have shown that the rolling resistance increases with the thickness of water on the road. This increase is approximately $30\%$ with light rain conditions of $0.3$mm of water on the road surface\cite{Wet_Rolling}. Table \ref{tab:rain} displays the time results of optimizations for different rain conditions, and the results when the actual rain conditions do not match those optimized for. Rain, by increasing rolling resistance (making it difficult to go fast on straightaways) and decreasing static friction (forcing cyclists to slow down more around corners) has a universally negative effect under our model. Yet we see that, by optimizing with consideration of the weather conditions, its effects can be mitigated to the point that the stochastic optimization produces a slightly better finishing time for rain than for no rain.

\begin{figure}
    \centering
    \includegraphics[width=0.5\textwidth]{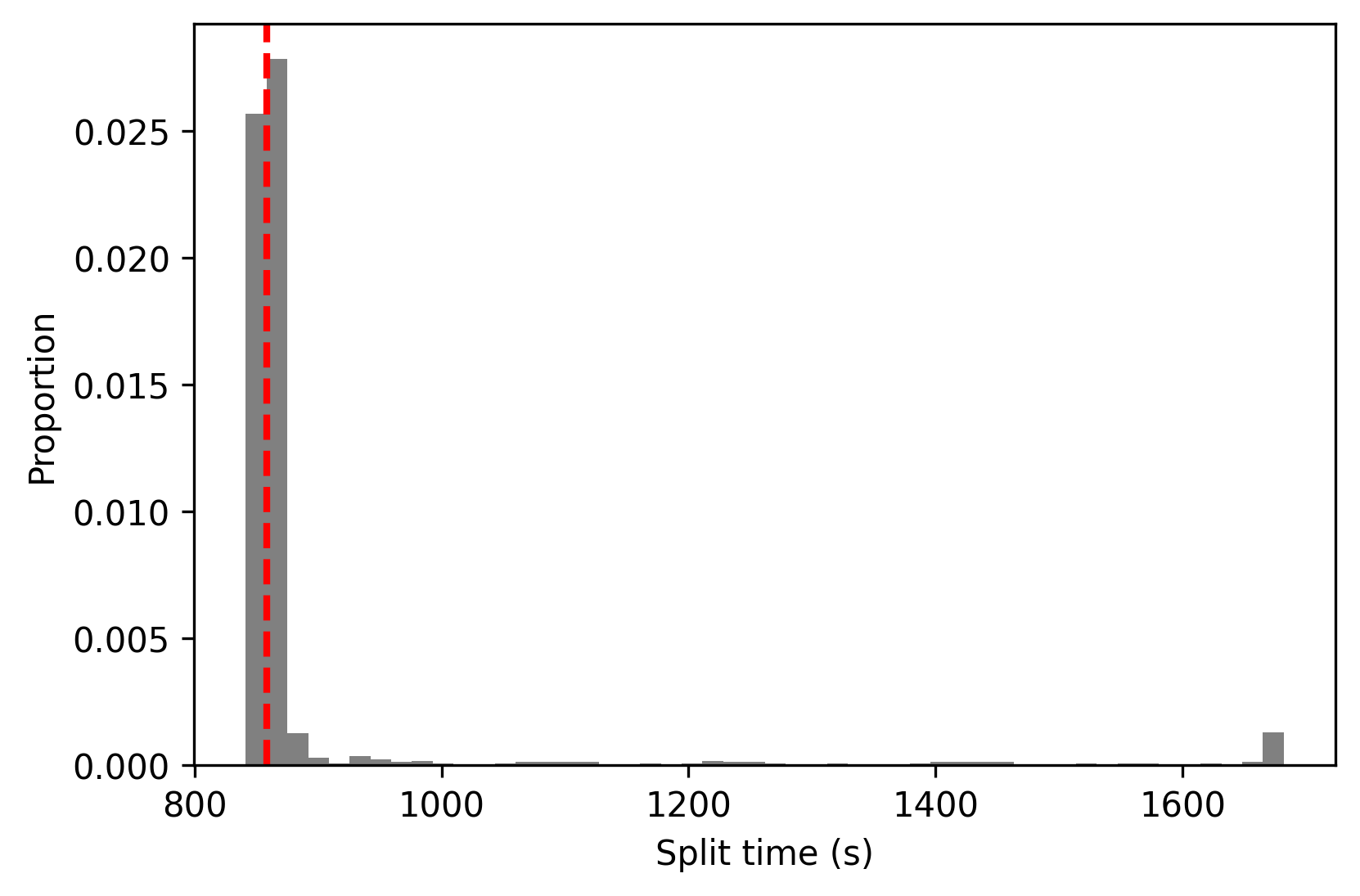}
    \caption{Filippo Ganna's split times for the split from to 25,753m to 32,360m in 1000 simulations with deviations from the optimal power strategy. In every segment, the deviation from the optimal power level $P$ is calculated under a Normal distribution with a mean of zero and a standard deviation of $P/50$. Under the optimal strategy, the split time is 858.5s (indicated with the red dashed line). Note that most split times fall around the optimal time. However, some are significantly slower. These are likely choices for power levels which make Ganna exhausted by the time this split is reached.}
    \label{fig:split}
\end{figure}
\subsection{Sensitivity to Strategy}
Given a power strategy, the expected split times and deviations from them may be computed. One split of interest in the Tokyo 2020 men's time trial race is the time it takes to climb the slope that runs from about 26,000m to 32,000m. Figure \ref{fig:split} displays Filippo Ganna's split time for this split when deviations from the optimal strategy are made. We see that the model is indeed sensitive to how closely the cyclist follows the strategy, as the optimization naturally creates strategies which leave energy at 0 by the end of the time trial. If a cyclist exceeds the target power too much, they will be fatigued and must bike at their baseline speed.

\begin{figure}[ht]
    \includegraphics[width=0.28\textwidth]{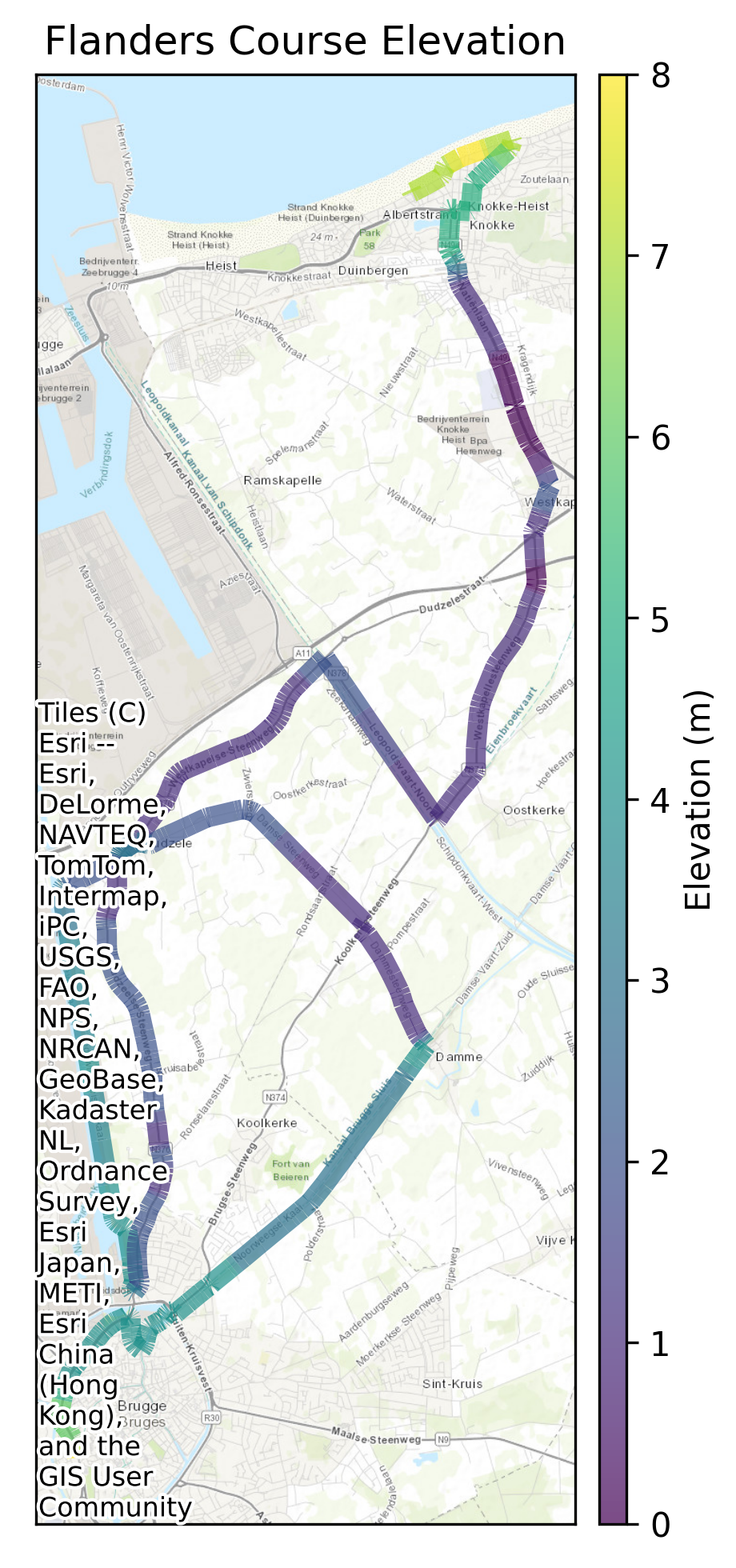}
    \centering
    \includegraphics[width=0.71\textwidth]{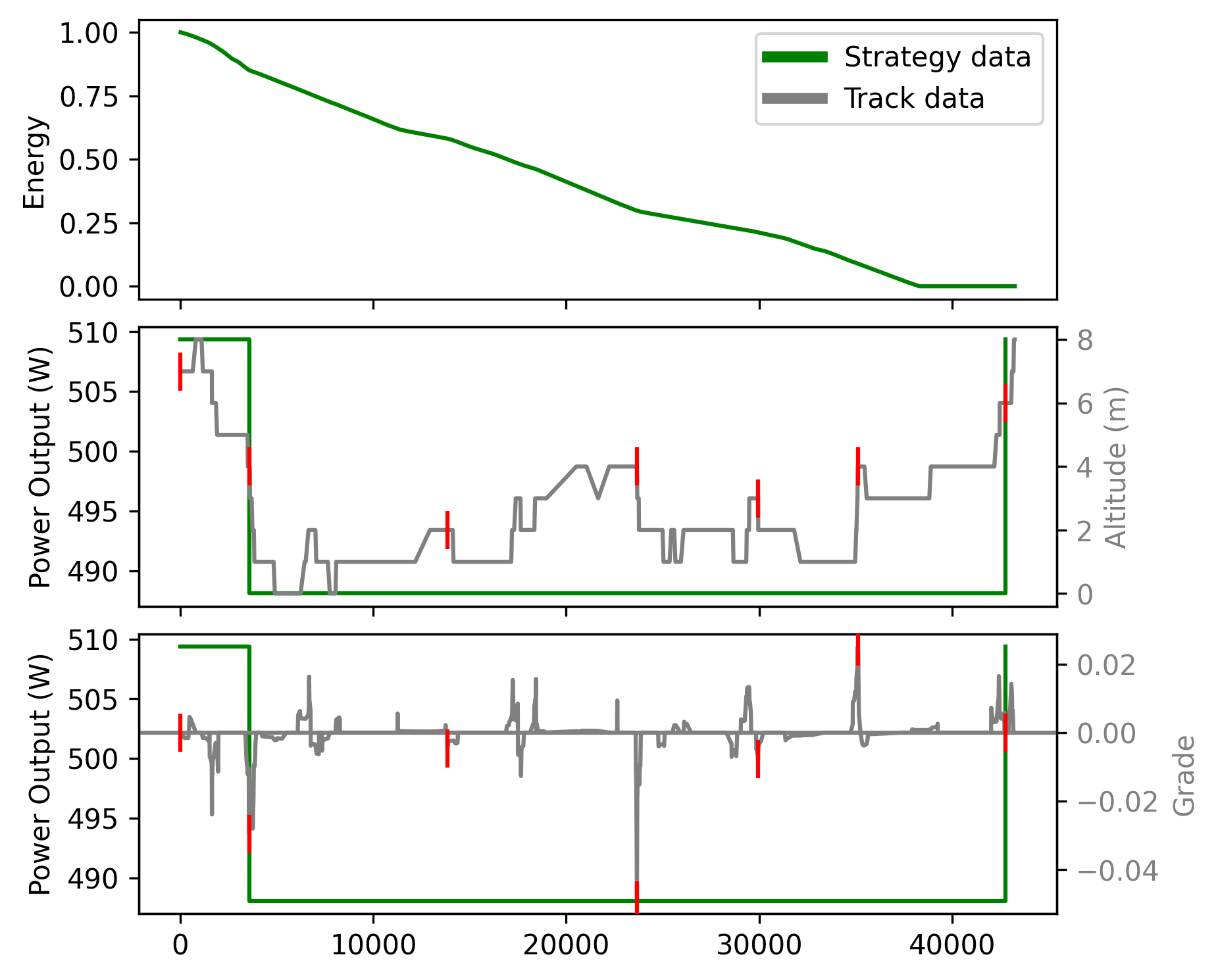}
    \caption{Filipo Ganna's power curve in Flanders, Belgium. This track has him mostly riding at the same power, and this is due to the track being mostly flat and a straight course without many curves. This lends itself to an optimal strategy where constant power (though with a burst of power at the beginning) leads to a better time.}
    \label{fig:flanders}
\end{figure}

\begin{figure}[ht]
    \includegraphics[width=0.39\textwidth]{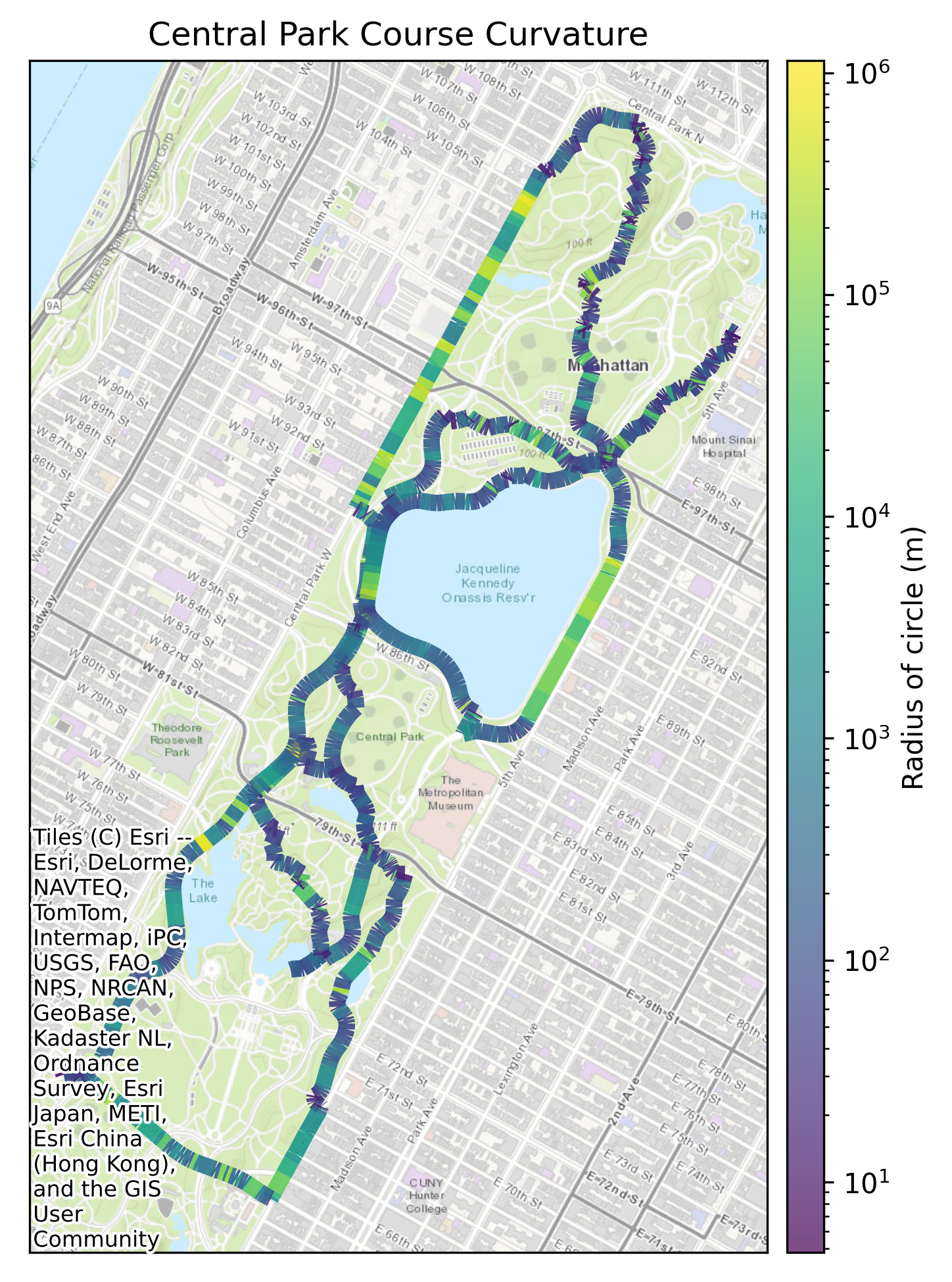}
    \includegraphics[width=0.6\textwidth]{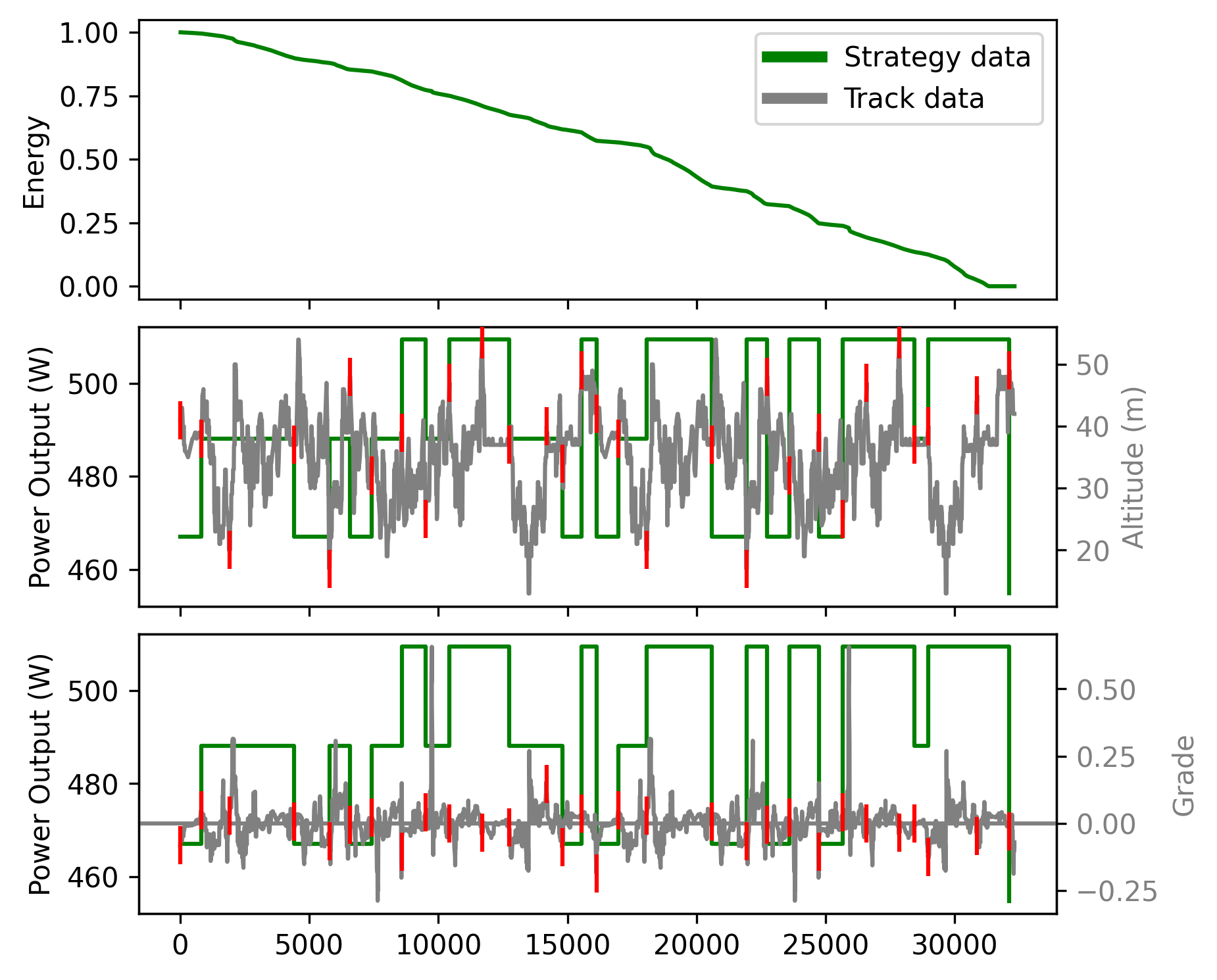}
    \caption{Filipo Ganna's power curve in a custom designed map of NYC Central Park. This map is designed to go around the bike path across all of the park, and will have sharp curves as the reach the edges. Along the path there are various short hills where he is seen changing his power output to match the slope. }
    \label{fig:NYC}
\end{figure}

\subsection{Results for other courses}
Since these strategies are tailored to the rider and we are able to test out multiple riders on the same Tokyo track, the next benchmark is being able to place the same rider on different tracks and have them optimize their power for each track respectively. The results below in Figure \ref{fig:flanders} show Filippo Ganna's optimal racing plan for the UCI 2021 World Championships track in Flanders. The drastic difference in racing plan is proof that the rider is also optimizing their plan to tackle the unique aspects of each individual race, but also that this model is general enough to accommodate different riders and different courses.

\section{Model Evaluation}
\subsection{Overall Model}
\paragraph{Strengths}
\begin{itemize}
    \item The model gives results that roughly coincide with real-world results. For instance, Ganna finished in 56 minutes in the Tokyo 2020 time trial \cite{56min}, which is reasonably close to his finishing times when weather matches that expected in Table \ref{tab:rain} (around 68 minutes). This goes to show that our models for power curves, fatigue, and kinematics are roughly correct.
    \item Using real-world data for routes and power curves makes the model directly applicable to specific races and athletes.
\end{itemize}
\paragraph{Limitations}
\begin{itemize}
    \item For weather and rain, it assumes that these effects are constant throughout the race. In reality, wind can change direction and rain will either increase or decrease over time. Since our model returns a power based on the position, it would be hard to constantly update weather especially when the rider might deviate from split times.
    \item The MMP at certain intervals are from race and practice data, which might not accurately reflect a cyclist's power curve as well as laboratory tests would show.
\end{itemize}

\subsection{Differential Equation Model}
\paragraph{Strengths}
\begin{itemize}
    \item The kinematics model is very well-behaved. For instance, in Figure \ref{fig:wind-finish}, the finishing times change smoothly as windspeed and heading change.
    \item The model runs in $O(\sqrt{n})$, where $n$ is the number of simulations done.
\end{itemize}
\paragraph{Limitations}
\begin{itemize}
    \item As the forwards force from the cyclist's power is determined by $P(x)/v$, the model is numerically unstable when $v$ is small. This is solved by integrating with sufficiently small timesteps.
\end{itemize}

\subsection{Fatigue Differential Equation Model}
\paragraph{Strengths}
\begin{itemize}
    \item The model is simple and intuitively matches what a power curve represents.
    \item The rate at which energy decreases is directly determined by the power curve, and not through parameters that roughly summarize a power curve.
\end{itemize}

\paragraph{Limitations}
\begin{itemize}
    \item We assume a linear rate of recovery in relation to power for power levels below critical power, which is most likely not entirely accurate
\end{itemize}

\subsection{Tree Exploration with Monte-Carlo Evaluation}
\paragraph{Strengths}
\begin{itemize}
    \item The optimization algorithm is able to search a very large search space efficiently. Notably, it is linear in the number of segments the track is split into, as well as the number of power levels evaluated for each segment. It thus perfectly matches the optimization problem at hand.
    \item The algorithm takes advantage of the differential equation model's ability, when vectorized, to efficiently solve for the finishing times of many power strategies at once.
    \item The algorithm quickly converges to an optimal solution. For example, changing $n$ from 100 to 1,000 only improves the solution by 23 seconds.
\end{itemize}
\paragraph{Limitations}
\begin{itemize}
    \item As optimization is only done over discrete power levels, the model may miss better power strategies which take on intermediate power levels.
    \item Similarly, as optimization is only done over discrete segments of the track, the model may miss better power strategies which continuously vary power output over the whole track.
    \item As noted above, the optimal strategy runs cyclists' energy down to zero by the end of a race, which is obvious as, if any energy is remaining by the end, the cyclist could have outputted more power and gone faster earlier. Any deviations from the plan could cause the cyclist to be completely fatigued by the end of the race. As the optimization algorithm takes the $k$th best time for a choice as the time to evaluate a node with, it cannot naturally take into account randomness in execution like traditional Monte Carlo Tree Search. In practice, during a race, a cyclist could try to err on the side of caution, and leave some energy to the end, and indeed we often see cyclists using up extra energy in a sprint at the end of a time trial.
\end{itemize}

\section{Discussion of Multi-Rider Team Trials}
Another component of cycling competitions are the team time trials, where 6 riders compete in a team and are judged by the finishing positions of the first 4 in the group. This leads to alternative strategies where riders optimize for the team rather than individual results, and can create room for specialized cyclists such as the Domestique \cite{Domestique}, who leads the team at the beginning and allows them to conserve energy by taking on all of the wind resistance. 

Optimal results can be simulated by including the power curves of each rider in the team, and optimizing for the finish speed of any 4 riders. With the assumption that the riders will be in a linked formation, only the front rider will experience the effects of wind resistance, and the rest can cycle in the slipstream generated. One example would be allowing the Domestiques to exceed their power at the start and run out of energy, if it allowed the other four riders to break out and sprint towards the finish line.

Different strategies that can be simulated include rotating the lead position between the riders, which shares the burden of wind resistance equally across all riders. Since the model optimizes for minimizing time, by including a choice to rotate positions at each segment, the model will also optimize the rotation between riders.

By running the optimization over a large set of riders, it could also find the optimal 6 riders to fit within a team, based on how their combination optimizes the minimum time.

\section{Conclusion}
Using the innovative Tree Exploration with Monte-Carlo Evaluation algorithm, we were able to provide a race plan personalized for the rider and the specific track they are competing on. At any point on the course, the rider will know how much power to output, as well as their expected times for reaching different checkpoints and finishing the race overall. 

This model incorporates information about the rider's power levels, weight, and individual stamina calculated from the Omni-PD model. Data for the tracks is precise, and allows the model to consider terrain conditions at each specific point in the race, rather than merely generalizing from similar tracks. Using the differential equation model, these factors at instantaneous positions along the track are able to be translated into instantaneous rates of change in state conditions, namely cyclist energy levels, distance covered along the track, and velocity. By modeling these incremental changes over time, we are able to get a very accurate idea of the performance of individual power strategies, providing a framework for the optimization process.

Also included as inputs are race-day wind direction and weather conditions, allowing the Directeur Sportif to prepare for all possibilities during the competition. The plan comes with expected split times at each segment, and a range of deviations depending on how closely the plan was followed. This is useful, as riders realistically will differ slightly from the optimal plan, but will still be able to get an estimation of their lap time. The resulting race plan is promising, due to its incorporation of real terrain and physiological factors, robustness to deviations, and ease of access to cyclists.




\addcontentsline{toc}{section}{References}
\bibliographystyle{amsplain}
\small
\bibliography{main} 


\normalsize
\setcounter{secnumdepth}{-1}
\newpage
\section{Letter to Directeur Sportif}

Dear Directeur Sportif,\\

    \noindent As you may very well know, on race day mere seconds differentiate the cyclist who comes out on top from those who don't even place. As athletes near the limits of human potential, utilizing data regarding individual cyclist capabilities as well as the track on which they'll compete will become increasingly necessary in order to devise personalized strategies to gain a competitive edge. Metrics for cyclist performance are getting evermore precise, and hence it's only natural that the way it is utilized to optimize race strategy keeps up. 
    
    The model takes in the specific characteristics of a cyclist and the track to create a racing plan tailored specifically for them. This way, their individual strengths are utilized to the utmost, and the rider can be more confident in the plan.
    
    This is done by creating a representation of the cyclist, and repeatedly simulating them racing the course thousands of times. At each key part of the race, they are able to decide how much power they will cycle with for the next segment. Simulations of the possible finishing times for each choice are done, and the power level that leads to the best time is selected. The fastest time for the course is then saved, along with the power the cyclist needs to expend at each part of the race to achieve that time. The result is a racing plan that is effective and simple to follow.
    
    Attached to this letter is the map of next year's 2022 UCI World Championship Men's Individual Time Trial event at Wollongong, Australia. Next to it is a race plan we have designed specifically for Filippo Ganna, the time trial specialist, on this race track. 
    
    \begin{figure}[ht]
    \includegraphics[width=0.39\textwidth]{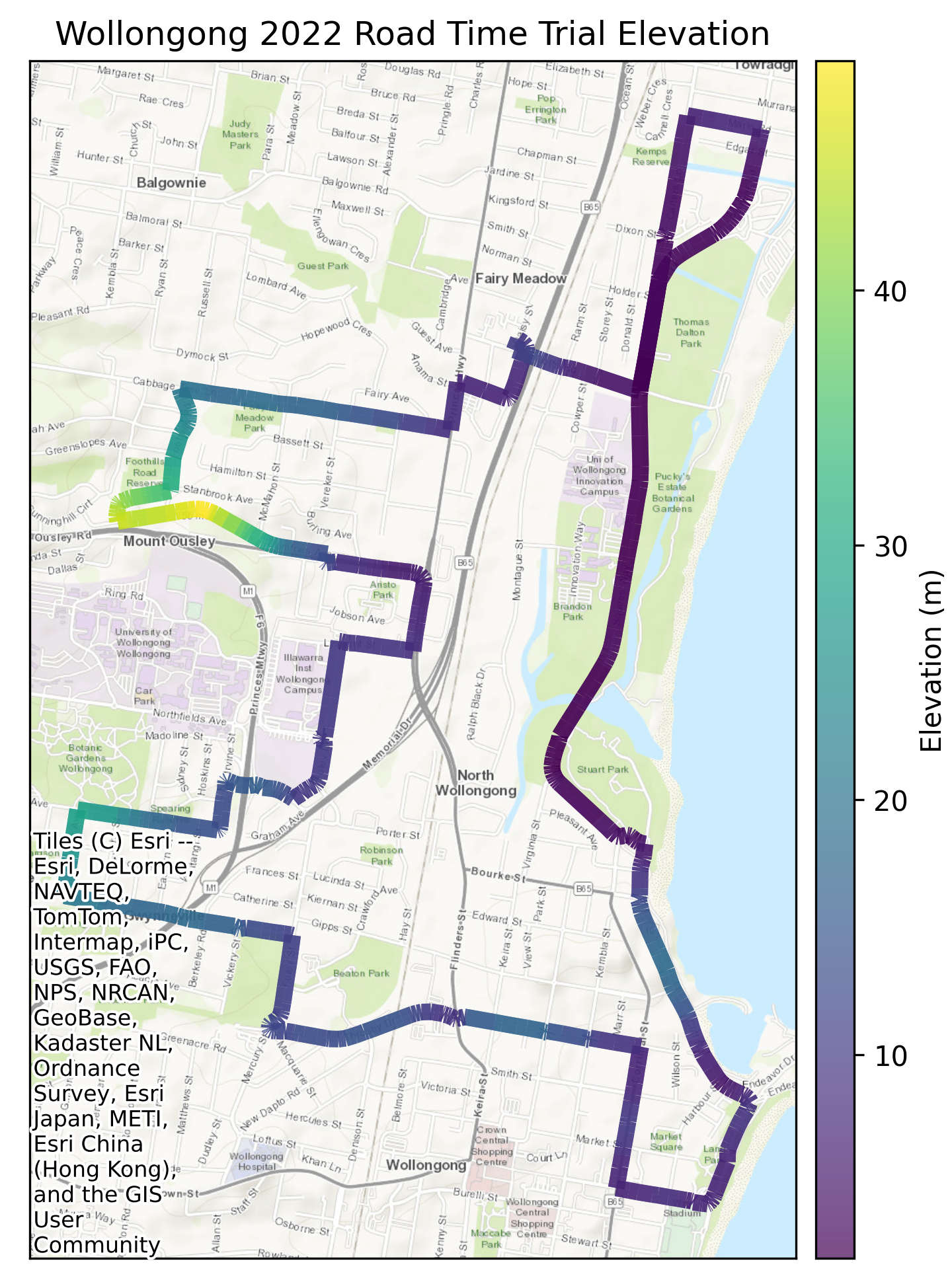}
    \includegraphics[width=0.59\textwidth]{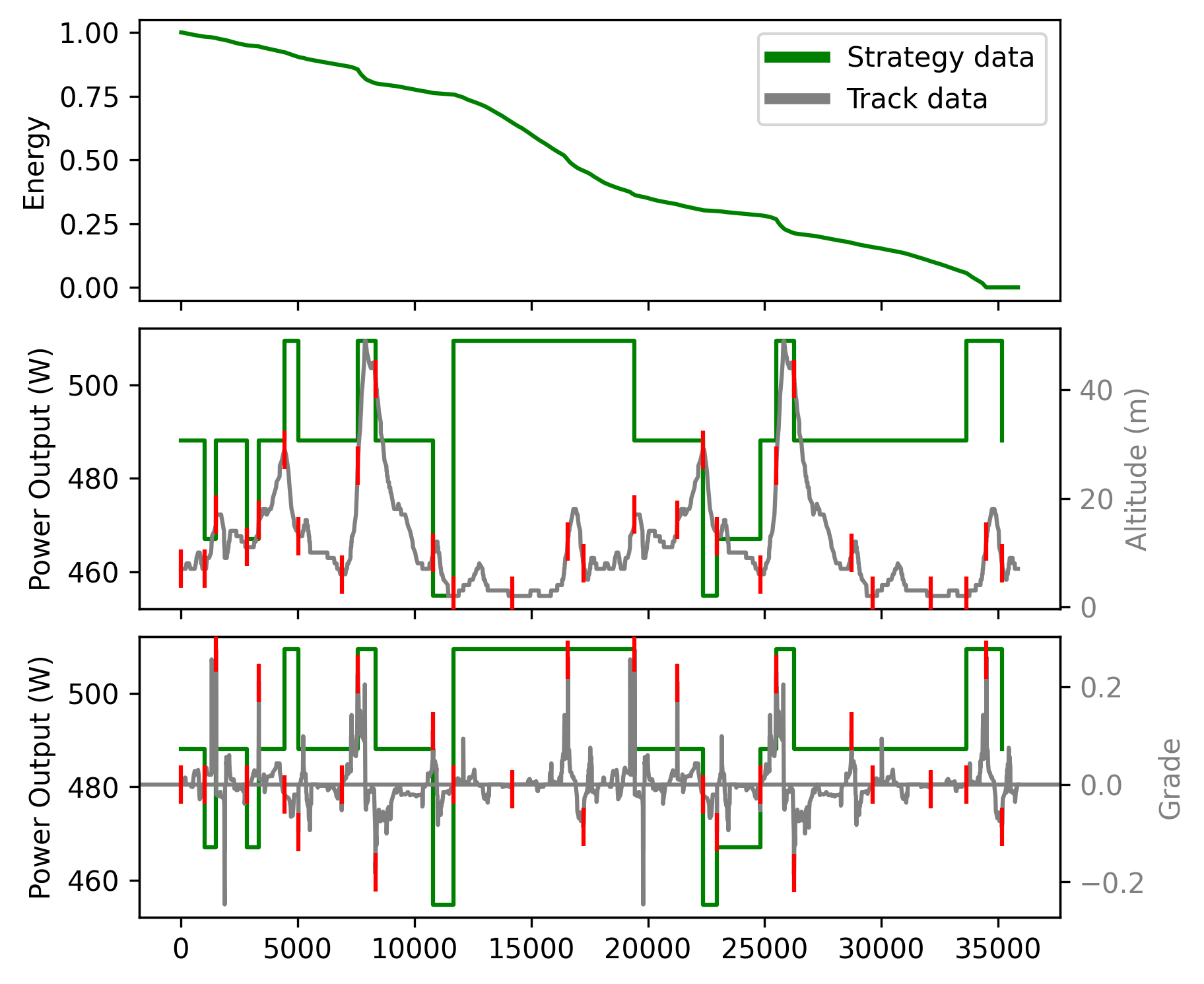}
    \caption{UCI 2022 Cycling track and Filippo Ganna's optimal strategy. Under this strategy, he can finish the race in 51 minutes}
    \end{figure}
    
    You can see that in the middle chart that Filippo's power output increases for climbs up hills and decreases on descents, which matches common intuition. The strength of this model lies in the precision of how much power should be used at each segment. On the top is also a chart of his energy over time, which decreases accordingly to how hard he cycles in each part. The power output is segmented by key parts of the race, so that it is easier for the rider to follow the plan, and at any point he can quickly glance at exactly how much power he needs.
    
    We also account for the maximum speed he can take at turns, and the acceleration he gets going downhill from gravity, to ensure his safety.
    
    \begin{figure}[ht]
        \begin{center}
            \includegraphics[width=0.59\textwidth]{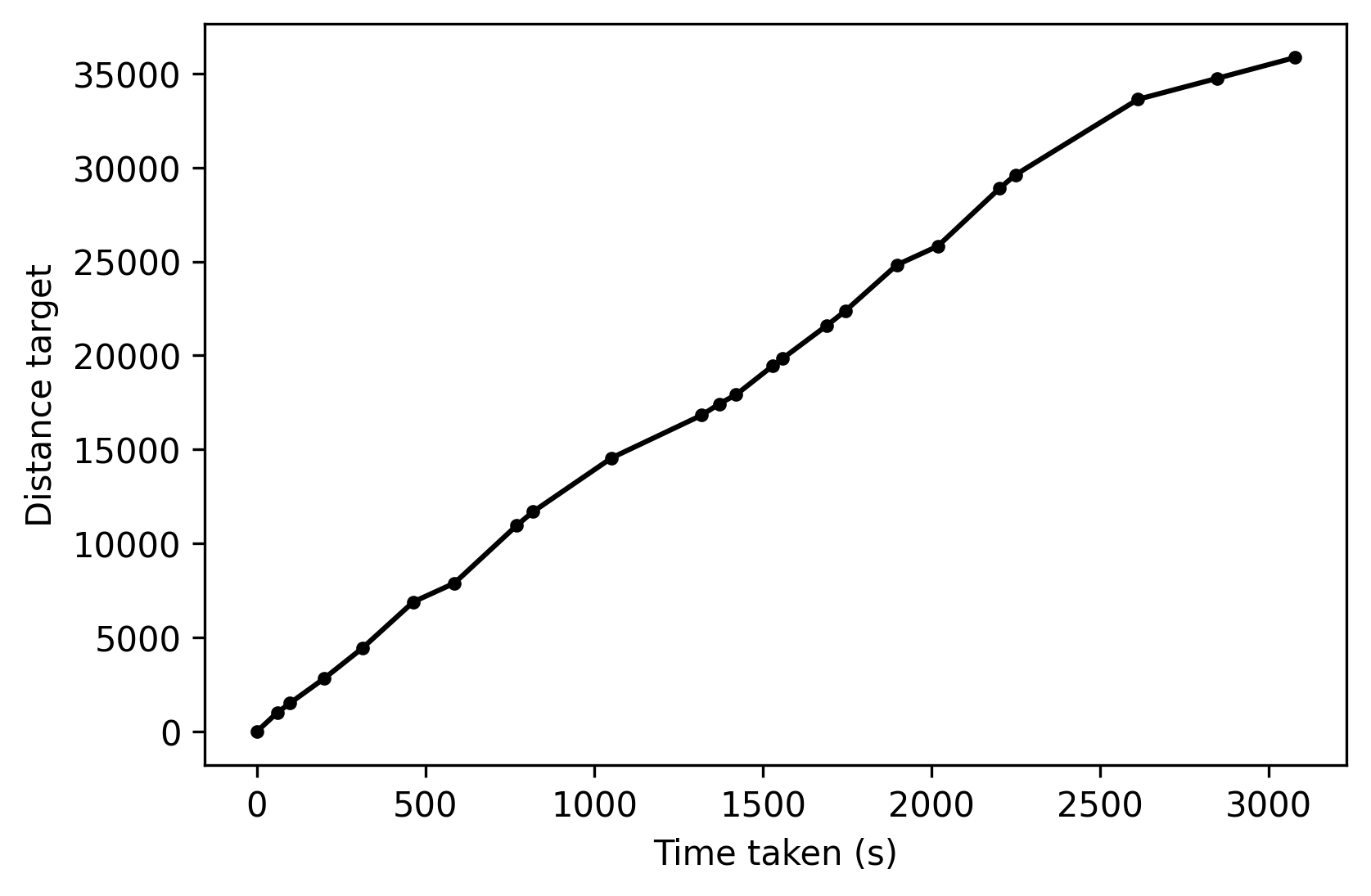}
            \caption{Distance of the track plotted against the time he will take to reach each split.}
        \end{center}

    \end{figure}
    
    As part of the model, we have split times for each of the key segments. The total projected time he will take to finish the course is 51 minutes. These numbers can be used to track his progress along the race and to see if he is above or below target. One example is knowing that he will reach the peak of the first large hill 9:45 mins into the race, and that at that point he should be putting in more than 500 watts of power. The model is able to compute deviations from this, in case he misses the target, and can give an expected range of when he will reach each segment. 
    
    As the event gets closer, you get a better sense of the weather conditions on race-day. These can be factored into the calculations by giving the model expected wind speed, wind direction, and rain. All of this will ensure that you will get the most realistic model possible and the best strategy for Filippo.\\

\noindent Best Regards,\\
The Bicycling Optimization Interest Society

\end{document}